\documentclass[]{aastex631}
\usepackage[latin9]{inputenc}
\makeatletter

\usepackage{amsmath}
\usepackage{mathtools}
\newcommand\numberthis{\addtocounter{equation}{1}\tag{\theequation}}
\usepackage[normalem]{ulem}
\usepackage{soul}
\usepackage{hyperref}
\newcommand{\be}{\begin{eqnarray}}

\newcommand{\ee}{\end{eqnarray}}

\makeatother

\newcommand{\aeg}{a^{(\rm EG)}}
\newcommand{\amw}{a^{(\rm MW)}}

\begin{document}

\title{Probing the global 21-cm background by velocity-induced dipole and
quadrupole anisotropies }

\author{Selim~C.~Hotinli}
\email{shotinl1@jhu.edu}
\affiliation{\jhu}

\newcommand{\jhu}{William H. Miller III Department of Physics and Astronomy, Johns Hopkins University, Baltimore, MD 21218, USA}

\author{Kyungjin Ahn}
\affiliation{Department of Earth Sciences, Chosun University, Gwangju 61452, Korea}
\email{kjahn@chosun.ac.kr; corresponding author}

\begin{abstract}
The motion of an observer in the rest frame of the cosmic 21-cm background induces an anisotropy in the observed background, even when the background is isotropic. The induced anisotropy includes a dipole and a quadrupole, in the order decreasing in amplitude. If observed, these multipole anisotropies can be used as additional probes of the spectral shape of the global 21-cm background for mitigating the ambiguity in the monopole spectrum probed by single-element radio telescopes such as EDGES and SARAS. This could also help with understanding the astrophysical and cosmological processes that occurred during the cosmic dawn and the epoch of reionization, and even improving on the estimation of the solar velocity and the foreground spectra. Here, we study the feasibility of such observations and present science drivers for the measurement of the 21-cm dipole and quadrupole. We find future 21-cm experiments can in principle detect the 21-cm dipole to high significance, potentially improving measurement accuracy of Earth velocity with respect to Milky-Way, galactic and extra-galactic foreground monopole spectra by an order of magnitude, as well as improving 21-cm astrophysical parameters by 2-5 percent.
\end{abstract}


\section{Introduction\label{sec:Intro}}

The measurement of the average absorption or emission of the cosmological redshifted 21-cm photons across the entire sky, or the global 21-cm signal, provides a clear window into the astrophysics of the early Universe. The global 21-cm signal is determined by the amount of neutral hydrogen present and by the relative number of electrons in the ground and excited states of the 21-cm transition, which is often defined in terms of the spin temperature of the transition. The spin temperature depends in turn on the kinetic temperature of the gas, and on the efficiency with which spin temperature can be driven towards the kinetic temperature, and away from the CMB temperature, by collisions and by Lyman-$\alpha$ coupling (the Wouthuysen-Field effect: \citealt{Wouthuysen1952,Field1958}). The global 21-cm signal is therefore sensitive to the evolution of a variety of forms of radiation in the early Universe, including: ionization radiation which destroys natural hydrogen, $X$-rays which heat the gas and raise the kinetic temperature, and the Lyman-$\alpha$ radiation which mostly comes from the redshifted soft UV photons falling on Lyman resonance lines. Therefore, the global 21-cm signal is expected to shed light on the appearance and evolution of the first stars, galaxies, and black holes in the Universe, and even new physics beyond the standard model, during the Dark Ages, the Cosmic Dawn, and the Epoch of Reionization.

A wide selection of experiments are aiming to measure the global 21-cm signal, such as EDGES~\citep{Bowman:2018yin}, SARAS~\citep{SARAS2022}, PRIZM~\citep{2019JAI.....850004P}, DAPPER~\citep{DAPPER2021} and LuSEE-Night~\citep{2023arXiv230110345B}. The main limitation faced by these experiments are the radio foregrounds such as the Galactic synchrotron emission, which dominate the global 21-cm signal in the radio bands by orders of magnitude, which needs to be subtracted or accounted for by modelling. The smoothness of the foreground spectra in frequency suggests that joint modelling of the large foregrounds simultaneously when probing the cosmology and astrophysics of reionization may indeed be possible. Similar to the cosmic microwave background (CMB), the 21-cm global signal is also subject to a higher-order relativistic corrections due to Earth's motion against the rest frame of the background (see e.g. \citealt{Bottani1992} and \citealt{Kamionkowski2003} for kinematically induced dipole and quadrupole moments of an isotropic background; \citealt{Challinor2002} for the first formulation of the relativistic aberration of general multipoles). Here, we calculate the kinematically induced dipole and quadrupole anisotropies and assess the effect of these to our understanding of the 21-cm signal. Given the smallness of the signals compared to foregrounds, we find that it could be important to accurately account for the relativistic dipole and quadrupole effects when modelling the 21-cm global signal.

The dipole due to our motion has been measured with increasing accuracy, particularly with the \textit{Planck} satellite. The velocity has been found to be $v=(369.82\pm0.11){\rm km/s}$ in the direction $(l,b)=(264^\circ\!\!.021\pm0^\circ\!\!.011, 48^\circ\!\!.253\pm0^\circ\!\!.005)$. The CMB dipole is too large to be explained as the intrinsic one in the standard model, and thus this velocity is deduced with the assumption that the CMB dipole has the kinematic origin only. Whether there exists an intrinsic dipole that is comparable to the kinematic dipole is under ongoing investigation \citep{KumarAluri2023} using not only the CMB \citep{Ferreira2021} but also the flux-limited distribution of lower-redshift objects such as quasars \citep{Secrest2021} and radio galaxies \citep{Xia2010,Singal2011,Chen2016}, leading to conflicting results (e.g. \citealt{Ferreira2021} for purely kinematic origin; \citealt{Secrest2021}, \citealt{Xia2010}, \citealt{Singal2011}, and \citealt{Chen2016} for partly intrinsic origin)\footnote{See also \citet{Balashev:2015lla,CORE:2017krr} on discussions about the spectral distortions as well as the extra-galactic foreground dipole from our motion.}. Here, we quantify the dipole, quadrupole and aberration effects induced by this velocity in the 21-cm background, under the assumption of the standard theory that the large-scale (a few lowest multipoles) 21-cm background also bears a near-perfect isotropy in its rest frame.

An important merit of measuring the induced 21-cm dipole and quadrupole is having redundancy on the spectral shape of the 21cm monopole signal. As will be shown in this paper, spectral shapes of these induced multipoles are solely determined by that of the 21-cm monopole (see \citealt{Deshpande2018} for using the diurnal modulation of the 21-cm dipole to probe the monopole spectrum). This is important because the estimated 21-cm monopole spectrum so far is still debatable. The 21-cm signal measured by EDGES and analyzed by the EDGES team show $\sim 500$~mK absorption spectrum and cannot be easily explained in the $\Lambda$CDM cosmology. This stimulated scenarios beyond the standard model (e.g. \citealt{Tashiro2014,Barkana2018,Hill2018}) or more mundane but somewhat unorthodox models for either the foreground (e.g. \citealt{Draine2018}) or the background (e.g. \citealt{Mirocha2019}). In contrast, data measured by SARAS~3 refutes the claim by the EDGES team at $\sim 1.6\,\sigma$, showing consistency with the instrumental noise \citep{SARAS2022}. While the result of SARAS~3 may be taken as a null experiment to constrain the high-redshift astrophysics significantly \citep{Bevins2022}, it is still difficult to make a firm conclusion due to the impact of the beam chromaticity \citep{SARAS2022}, the ionosphere \citep{Shen2021}, and other systematic uncertainties {\citep{Hills2018,Bradley2019,Singh2019,Spinelli2019,Sims2020,Bevins2021,SARAS2022}}. There is even a possibility that the global 21-cm background spectrum is smooth during the Cosmic Dawn, even when the star formation activity is significant: the predicted global 21-cm signal of the SRII model category in \citet{Ahn2021} is spectrally featureless at $z\gtrsim 16$ due to self-regulated formation of Population III stars. Such a featureless 21-cm signal is likely to be removed together with the foreground and mimic a null global 21-cm signal. Therefore, a redundant measurement of the monopole spectrum can provide a crucial resolution to settle these issues.

This paper is organized as follows. Section~\ref{sec:formal} is devoted to quantifying the low-order multipole moments of the 21-cm background and the foreground. In Section~\ref{sec:induced}, we calculate the Doppler effect of and the relativistic corrections to the 21-cm global signal spectrum leading to the kinematically induced multipole spectra. In Section~\ref{sec:intrinsic}, the intrinsic multipoles of the foreground are defined. 
In Sections~\ref{sec:FGinduced} and \ref{sec:observed}, we quantify how the kinematically induced dipole and quadrupole moments from the intrinsic foreground multipoles add to the 21-cm background.
In Section~\ref{sec:21cmfast} we show how we generate a fiducial reionization model that will be used for error estimates.
In Section~\ref{sec:21cmfast} we describe the foreground monopole and our fiducial 21-cm signal. In Section~\ref{sec:Forecasts}, we describe the results of our Fisher analysis with focused interest in observational forecasts on parameters for the reionization model and other factors including the solar velocity and the foreground.  Summary and discussion are in Section~\ref{sec:discussion}. Some details left out in the main text are presented in the Appendix.

\section{Low-order multipole moments of the 21-cm background and the foreground\label{sec:formal}}
\subsection{Kinematically induced dipole and quadrupole of the 21-cm background\label{sec:induced}}

We follow \citet{Bottani1992} for the base equations to start with.
The observed frequency (temperature) $\nu'$ ($T'$)
of a background is related to the source frequency (temperature) $\nu$
($T$) as
\begin{equation}
\frac{\nu'}{\nu}=\frac{T'(\nu')}{T(\nu)}=\gamma(1+\beta\mu)=\frac{1}{\gamma(1-\beta\mu')},
\label{eq:nu}
\end{equation}
where $\beta=v/c$ with the velocity of an observer against the background
rest frame $v$ and the speed of light $c$, $\gamma=(1-\beta^{2})^{-1/2}$,
and $\mu'\equiv\cos\theta'$ with the angle $\theta'$
between the line of sight and the direction of $v$ \citep{Peebles1968,Bottani1992},
such that the point on the sky the observer is moving toward has $\theta'=0$.
Variables with the superscript ``$'$'' and without are
those measured in the observer rest frame and the background (source)
rest frame, respectively. The last equality in equation (\ref{eq:nu}) is due to the  aberration of light:
\begin{equation}
\mu'=\frac{\mu+\beta}{1+\beta\mu},
\label{eq:mu}
\end{equation}
with $\mu\equiv\cos\theta$. Using the Liouville's invariance theorem
for the specific intensity $I(\nu)$, $I'(\nu')\nu'^{-3}=I(\nu)\nu^{-3}$,
and expanding $I'(\nu')$ in terms of $I$ and $\nu'$,
the measured specific intensity by the moving observer becomes\footnote{The original equation in \citet{Bottani1992} has a typographical error, having $(1/2)(3-\alpha)\beta^2$ in the equivalent of our equation (\ref{eq:Inu}). Nevertheless, because this term affects only the monopole at $\mathcal{O}(\beta^2)$ and is approximated to zero, the scientific result of \citet{Bottani1992} is intact.}
\begin{equation}
I'(\nu')=I(\nu')\left\{ 1+\frac{1}{2}({\alpha-3})\beta^{2}+(3-\alpha)\beta\mu'+\left[6-3\alpha+\frac{1}{2}\frac{\nu'^{2}}{I(\nu')}I^{\prime\prime}(\nu')\right]\beta^{2}\mu'^{2}\right\} ,
\label{eq:Inu}
\end{equation}
where $I$ is the monopole measured at the background rest frame,
$I^{\prime\prime}\equiv d^{2}I(\nu)/d\nu^{2}$ and $\alpha\equiv d\ln I(\nu)/d\ln\nu$)
is the spectral index of $I$ at given $\nu$. Due to the obvious azimuthal symmetry, the angle dependence is omitted in the expression: $I'(\nu')\equiv  I'(\mu',\,\nu')$.

We now focus on the 21-cm background. Because we are interested in the high-redshift regime, where the structure formation is still linear in large scale and reionization has barely started, it is a good approximation that the 21-cm background is isotropic and shares its rest frame with the CMB in the scale of interest.\footnote{Note, however, that these assumptions may fall short from matching observations with increasingly precise measurement of the cosmological dipoles and new developing methods~\cite[see~e.g.~][]{Cayuso:2021ljq}, as suggested by the long-existing `dipole tension'~\citep[see~e.g.][and references therein]{Dalang:2021ruy}.}. With this assumption, we start from the global intensity
measured in the background rest frame, $I$. Due to the interplay
between the 21-cm line and the CMB, the net intensity is given by
\begin{equation}
I(\nu)=I_{{\rm BB}}(\nu;\,T_{{\rm CMB,0}}){\rm e}^{-\tau(\nu)}+I_{{\rm BB}}(\nu_{21};\,T_{{\rm s}}(\nu))(1-{\rm e}^{-\tau(\nu)})(\nu/\nu_{21})^{3},
\label{eq:I21}
\end{equation}
where $I_{{\rm BB}}(\nu;\,T)\equiv(h_{{\rm P}}\nu^{3}/2c^{2})(\exp[h_{{\rm P}}\nu/kT]-1)^{-1}$
is the specific intensity of a black body with temperature $T$ measured
at frequency $\nu$, $h_{{\rm P}}$ is the Planck constant, $c$ is
the speed of light, $k$ is the Boltzmann constant, $T_{{\rm CMB,0}}=2.726\,{\rm K}$
is the present-day CMB temperature, $T_{{\rm s}}(\nu)$ is the spin
temperature, $\tau(\nu)$ is the global 21-cm optical depth, and the
observed frequency $\nu$ works as an implicit indicator of the cosmological
redshift $z$ such that $\nu=\nu_{21}/(1+z)$ with the 21-cm line
frequency $\nu_{21}=1.4204\,{\rm GHz}$ (e.g. \citealt{Morales2004}).
Both the CMB and the 21-cm emission background in the frequency range of our interest are deep in the Rayleigh-Jeans
regime when observed at redshifted 21-cm line frequencies, and thus
they do not contain the ``Wien quadrupole'' which comes not from
the relativistic aberration (equation \ref{eq:mu}) but from the sheer
mathematical conversion of the temperature into the specific intensity
in the Wien regime (\citealt{Bottani1992}). The optical depth is
approximated as an on-site one assuming a $\delta$-function line
profile as usual, given by
\begin{equation}
\tau(\nu)=0.01047x_{{\rm HI}}\left(\frac{\Omega_{b}h^{2}}{0.022239}\right)\left(\frac{1-Y}{0.753}\right)\left(\frac{h}{0.703}\right)^{-1}\left(\frac{E(z)}{19.217}\right)^{-1}\left(\frac{1+z}{11}\right)^{3}\left(\frac{T_{{\rm s}}(\nu)}{27.26\,{\rm K}}\right)^{-1},
\label{eq:tau}
\end{equation}
which lies in an optically thin regime, with the present-day baryon
abundance $\Omega_{b}$, the helium abundance $Y$, and the ratio
of the Hubble constant at past to now $E(z)\equiv H(z)/H_{0}$ (e.g.
\citealt{Morales2004}). We thus safely use the usual approximation
of equation (\ref{eq:I21}) in the Rayleigh-Jeans and the optically
thin regime:
\begin{equation}
I(\nu)\approx\left(\frac{kT_{{\rm CMB,0}}}{2c^{2}}\right)\nu^{2}\left[1+\left(\frac{T_{{\rm s},0}(\nu)}{T_{{\rm CMB,0}}}-1\right)\tau(\nu)\right]=\left(\frac{kT_{{\rm CMB,0}}}{2c^{2}}\right)\nu^{2}\left[1+\frac{\delta T_{b}(\nu)}{T_{{\rm CMB,0}}}\right],\label{eq:Inu_approx}
\end{equation}
which is expressed to clearly show the $\nu$-dependence, with the
spin temperature scaled to the present day, $T_{{\rm s},0}(\nu)\equiv T_{{\rm s}}(\nu)/(1+z)$,
and the differential brightness temperature $\delta T_{b}(\nu)\equiv(T_{{\rm s},0}(\nu)-T_{{\rm CMB,0}})\tau(\nu)$.
Regardless of the strength of $T_{{\rm s}}(\nu)$, even when $T_{{\rm s}}\gg T_{{\rm CMB}}$,
$\delta T_{b}(\nu)\ll T_{{\rm CMB,0}}$ because the global optical
depth $\tau(\nu)$ is small\footnote{The 21-cm optical depth of highly overdense regions can be large.
However, in this work, we are only interested in the global 21-cm
optical depth, which is small enough to be in the optically thin regime.}. 

The specific intensity observed by a moving observer will be given
by the combination of equations (\ref{eq:Inu}), (\ref{eq:tau}) and
(\ref{eq:Inu_approx}). This leads to 
\begin{equation}
I'(\nu')\approx I(\nu')\left\{ 1+\frac{1}{2}\left[{F(\nu')-1}\right]\beta^{2}+\left[1-F(\nu')\right]\beta\mu'+\left[1-F(\nu')+G(\nu')\right]\beta^{2}\mu'^{2}\right\} ,\label{eq:Inuo_approx}
\end{equation}
or in brightness temperature, thanks to the lack of the Wien quadrupole\footnote{One can also arrive at equation (\ref{eq:Tnuo_approx}) starting from the Liouville's invariance theorem for the brightness temperature, $T'(\nu')\nu'^{-1}=T(\nu)\nu^{-1}$.},
\begin{equation}
T'(\nu')\approx T(\nu')\left\{ 1+\frac{1}{2}\left[{F(\nu')-1}\right]\beta^{2}+\left[1-F(\nu')\right]\beta\mu'+\left[1-F(\nu')+G(\nu')\right]\beta^{2}\mu'^{2}\right\} ,\label{eq:Tnuo_approx}
\end{equation}
where the information of the differential 21-cm monopole, $\delta T_{b}(\nu')$,  is encoded in two functions $F$ and $G$:
\begin{align}
F(\nu') & \equiv \frac{1}{T_{{\rm CMB,0}}}\frac{d\delta T_{b}(\nu')}{d\ln\nu'},\nonumber \\
G(\nu') & \equiv \frac{1}{2}\frac{\nu'^{2}}{T_{{\rm CMB,0}}}\frac{d^{2}\delta T_{b}(\nu')}{d^{2}\nu'}.
\label{eq:FG}
\end{align}
One can also obtain the differential temperature (intensity), accurate to $\mathcal{O}(\beta^{2})$,
by subtracting the \emph{Doppler-shifted}, \emph{aberrated} and \emph{unattenuated} CMB:
\begin{align}
\delta T'(\nu') & =T'(\nu')-T_{{\rm CMB,0}}\left\{ 1-\frac{1}{2}\beta^{2}+\beta\mu'+\beta^{2}\mu'^{2}\right\} \nonumber \\
 & =T_{{\rm CMB,0}}\left\{ \frac{1}{2}F(\nu')\beta^{2}-F(\nu')\beta\mu'+\left[-F(\nu')+G(\nu')\right]\beta^{2}\mu'^{2}\right\} \nonumber \\
 & +\delta T_{b}(\nu')\left\{ 1+\frac{1}{2}\left[F(\nu')-1\right]\beta^{2}+\left[1-F(\nu')\right]\beta\mu'+\left[1-F(\nu')+G(\nu')\right]\beta^{2}\mu'^{2}\right\} \nonumber \\
 & \approx T_{{\rm CMB,0}}\left\{ \frac{1}{2}F(\nu')\beta^{2}-F(\nu')\beta\mu'+\left[-F(\nu')+G(\nu')\right]\beta^{2}\mu'^{2}\right\} \nonumber \\
 & +\delta T_{b}(\nu')\left\{ 1-\frac{1}{2}\beta^{2}+\beta\mu'+\beta^{2}\mu'^{2}\right\} \nonumber \\
 & =\delta T_{0}'(\nu')Y^{0}_{0}+\delta T_{1}'(\nu')Y^{0}_{1}(\mu')+\delta T_{2}'(\nu')Y^{0}_{2}(\mu'),
 \label{eq:dInuo}
\end{align}
where the approximation is allowed to keep only terms accurate
to $\mathcal{O}(\delta T_{b}/T_{{\rm CMB,0}})\sim F\sim G$, and the
final expression is the decomposition of $\delta T'(\nu')$
into the monopole $\delta T_{0}'$, the dipole $\delta T_{1}'$
and the quadrupole $\delta T_{2}'$ with the spherical harmonics coefficients 
$Y^{m=0}_{\ell}(\mu')$:
\begin{align}
\delta T_{0}'(\nu') & =\sqrt{4\pi}\left\{T_{{\rm CMB,0}}\left[\frac{1}{6}F(\nu')+\frac{1}{3}G(\nu')\right]\beta^{2}+\delta T_{b}(\nu')\left[1-\frac{1}{6}\beta^{2}\right]\right\}\approx\sqrt{4\pi}\delta T_{b}(\nu'),\nonumber \\
\delta T_{1}'(\nu') & =\beta\sqrt{\frac{4\pi}{3}}T_{{\rm CMB,0}}\left\{ -F(\nu')+\frac{\delta T_{b}(\nu')}{T_{{\rm CMB,0}}}\right\} ,\nonumber \\
\delta T_{2}'(\nu') & =\beta^{2}\sqrt{\frac{16\pi}{45}} T_{{\rm CMB,0}} \left\{ G(\nu')-F(\nu')+\frac{\delta T_{b}(\nu')}{T_{{\rm CMB,0}}}\right\} .\label{eq:multipoles}
\end{align}
Equation (\ref{eq:multipoles}) shows that, other than the 21-cm monopole
term, both $\delta T_{1}'(\nu')$ and $\delta T_{2}'(\nu')$
contain not only the derivatives of the monopole but also the monopole
itself. This is due to the Doppler shift for the dipole, and the light aberration for the quadrupole
(see e.g. equation (3) of \citealt{Bottani1992}). Note also that
neither the dipole nor the quadrupole has a dependence on $T_{{\rm CMB},0}$: since functions $F(\nu)$ and $G(\nu)$ that parameterize the differential 21-cm monopole defined in equation (9) are proportional {to} $1/T_{\rm CMB,0}$, $T_{{\rm CMB},0}$ cancels out from these expressions; and one can start from any foreground (e.g. inclusive of the
radiation from the Milky Way) for recalculating the dipole and the quadrupole.
At any rate, in principle, measuring $\delta T_{1}'(\nu')$
and $\delta T_{2}'(\nu')$ can give redundant information
in determining $\delta T_{b}(\nu')$. 

We note that subtraction of the CMB (equation \ref{eq:dInuo}) cannot be perfect, and uncertainties in measuring the (1) CMB temperature monopole $T_{\rm CMB,0}$ and (2) the solar velocity $\beta$ (amplitude and direction) can lead to uncertainties in $\delta T_{b}(\nu')$. We also note that the foreground can bear its own uncertainties. In Section \ref{sec:Forecasts}, we therefore use the net observed brightness temperature
\begin{equation}
T_{\rm net}(\hat{n}',\,\nu') \equiv T'^{\rm FG}(\hat{n}',\,\nu')+\Delta T_{\rm CMB,0}\left\{ 1-\frac{1}{2}\beta^{2}+\beta\mu'+\beta^{2}\mu'^{2}\right\}+\delta T'(\nu')
\label{eq:Ttotal}
\end{equation} 
composed of the foreground ${T'}^{\rm FG}$, the uncertainty caused by that of the CMB monopole, and the 21-cm monopole. Here, $\hat{n}'$ is the sky direction, $T_{\rm net}(\hat{n}',\,\nu')$ is the brightness temperature of the foreground, and $\Delta T_{\rm CMB,0}$ is the measurement uncertainty in $T_{\rm CMB,0}$. 

\subsection{Intrinsic multipole moments of the foreground\label{sec:intrinsic}}

The foreground, originating from the synchrotron, free-free emission and dust from the Milky Way, and also from {extra-galactic sources}, dominate the radio sky of our interest. Here we simply expand the observed total foreground brightness temperature at the sky direction $\hat{n}$ and the observing frequency $\nu'$ as
\begin{equation}
T'^{\rm (FG)}(\nu',\hat{n}')=\sum_{\ell,\,m}{a'}^{\rm (FG)}_{\ell m}(\nu') Y^{m}_{\ell}(\hat{n}') .
\label{eq:FGalm}
\end{equation}

Table~\ref{tab:coefficients} in the Appendix lists the dipole and quadrupole moments measured in a reference frame described below. The solar velocity against the background rest frame, ${\bf V}_{\odot,\,{\rm CMB}}$, defines $\hat{n}$ and $Y_{\ell}^{m}$ in a reference frame that is generated by rotating the galactic axes ($\hat{R}$, $\hat{\phi}$, $\hat{Z}$) with the Euler angles ($\alpha$, $\beta$, $\gamma$)=($l$, $\pi/2-b$, $0$) in the $z$-$y$-$z$ convention, where $l$ and $b$ are the galactic longitude and latitude of the vector ${\bf V}_{\odot,\,{\rm CMB}}$, respectively. The reference frame generated this way has $z$-axis pointing at $\hat{z} = {\bf V}_{\odot,\,{\rm CMB}}/V_{\odot,\,{\rm CMB}}$ and $y$-axis lying on the galactic plane. The recent CMB dipole observation gives $V_{\odot,\,{\rm CMB}}=369.82\pm 0.11$~km s$^{-1}$ and ($l$, $b$)=($264.02\pm 0.0085^{\circ}$, $48.253\pm 0.004^{\circ}$) (\citealt{2020A&A...644A.100P}; see also \citealt{Delouis:2021whm} for slightly different ${\bf V}_{\odot,\,{\rm CMB}}$).

\subsection{Induced dipole and quadrupole of the foreground\label{sec:FGinduced}}
It is important to fully understand the physical origins and properties of the dipole and quadrupole moments seen in the radio sky, in order to extract the 21-cm background component. The motion of the solar system against the foreground, dominated by the Milky Way, is expected to result in another set of induced moments. Indeed, the dipole and quadrupole of the low-frequency sky are dominated not only by the intrinsic multipoles of the foreground (see Section~\ref{sec:intrinsic}) but also by the kinematically induced ones of the foreground. The aberration and the Doppler shift induces, if intrinsic multipoles exist in the background rest frame, the recursive leakage of multipole moments (with $\ell$) into adjacent ones (with $\ell\pm 1$) at order $\mathcal{O}(\beta)$ in the observer frame. This has been first formulated and calculated up to $\mathcal{O}(\beta^2)$ by \citet{Challinor2002} (CvL2022 henceforth), and later calculated up to arbitrary order in $\beta$ by \citet{Chluba2011,Dai:2014swa}. Therefore, we need to quantify the leakage of multipole moments of the foreground into the dipole and quadrupole moments, which is a more demanding calculation than in the case of a presumably isotropic background in its rest frame.

We follow CvL2022 for calculating the induced multipoles of the foreground. If a foreground is observed in the foreground rest frame, the brightness temperature can be written as
\begin{equation}
T^{\rm (FG)}(\nu,\hat{n})=\sum_{\ell,\,m}a^{\rm (FG)}_{\ell m}(\nu) Y^{m}_{\ell}(\hat{n}).
\label{eq:FGalm_God}
\end{equation}

From the Liouville theorems $T'(\nu',\hat{n}'){\nu'}^{-1}=T(\nu,\hat{n}){\nu}^{-1}$ and $d\hat{n}'{\nu'}^{2}=d\hat{n}{\nu}^{2}$, and equations (\ref{eq:nu}), (\ref{eq:FGalm}) and (\ref{eq:FGalm_God}), one can obtain
\begin{equation}
{a'}_{\ell m}(\nu') = \sum_{l'}\int \frac{d\hat{n}}{\gamma(1+\beta \mu)} a_{\ell' m}(\nu) Y_{\ell' m}(\hat{n}) {Y}^{*}_{\ell m}(\hat{n}') = \sum_{l'} \mathcal{K}^{\ell'}_{\ell} a_{\ell' m}(\nu') ,
\label{eq:almtrans}
\end{equation}
where we dropped the subscript``(FG)'' for generality, and in both frames we set $\hat{z} = \hat{\beta}$ and also $\hat{x}$ and $\hat{y}$ do not vary. The notable difference of equation (\ref{eq:almtrans}) from equation (A1) of CvL2022 is that the ratio $\nu'/\nu$ is inverted because the former is for temperature while the latter is for specific intensity. Note also that the ``aberration kernel'' $\mathcal{K}^{\ell'}_{\ell}$ here is defined as a coefficient to $a_{\ell' m}(\nu')$ instead of $a_{\ell' m}(\nu)$, and thus requires Taylor-expanding $a_{\ell' m}(\nu)$ to $a_{\ell' m}(\nu')$. Now, up to $\mathcal{O}(\beta^2)$, the equivalent of  $\mathcal{K}^{\ell'}_{\ell}$ for the specific intensity is explicitly written in equation (A6) of CvL2022. For temperature multipoles, we simply need to replace every multipole $a_{\ell m}(\nu)$ with $a_{\ell m}(\nu)\nu^2$ in equation (A6) of CvL2022, or equivalently replace $\nu'(d/d\nu')$ and $\nu'^2(d^2/d\nu'^2)$ with $\nu'(d/d\nu')+2$ and $\nu'^2(d^2/d\nu'^2)+4\nu'(d/d\nu')+2$, respectively. Then, we obtain
\begin{align*}
\mathcal{K}^{\ell}_{\ell}=&
    1 + \beta^2 
    \left[
        \frac{1}{2} C^{2}_{(\ell+1) m} 
        \left(
            (\ell+1)(\ell+2)+2(\ell+2)\nu'\frac{d}{d\nu'}+\nu'^{2}\frac{d^2}{d\nu'^2}
        \right) 
        +\frac{1}{2} C^{2}_{\ell m}
        \left(
            \ell(\ell-1)-2(\ell-1)\nu'\frac{d}{d\nu'}+\nu'^{2}\frac{d^2}{d\nu'^2}
        \right) \right. \\
        & \left. +\frac{1}{2}
        \left(
            m^2-\ell(\ell+1)-1-\nu'\frac{d}{d\nu'}
        \right)
    \right] \\ 
\mathcal{K}^{\ell+1}_{\ell}=&-\beta C_{(\ell+1) m} \left[\ell+1+\nu'\frac{d}{d\nu'} \right], \\
\mathcal{K}^{\ell-1}_{\ell}=&-\beta C_{\ell m} \left[-\ell+\nu'\frac{d}{d\nu'} \right], \\
\mathcal{K}^{\ell+2}_{\ell}=&\frac{\beta^2}{2} C_{(\ell+2) m} C_{(\ell+1) m} 
    \left[
        (\ell+1)(\ell+2)+2(\ell+2)\nu'\frac{d}{d\nu'}+\nu'^{2}\frac{d^2}{d\nu'^2}
    \right], \\
\mathcal{K}^{\ell-2}_{\ell}=&\frac{\beta^2}{2} C_{\ell m} C_{(\ell-1) m}
    \left[
        \ell(\ell-1)-2(\ell-1)\nu'\frac{d}{d\nu'}+\nu'^{2}\frac{d^2}{d\nu'^2}
    \right], \\
\mathcal{K}^{\ell+3}_{\ell} \sim& \mathcal{K}^{\ell-3}_{\ell} \sim \mathcal{O}(\beta^3), \\ 
\vdots&    \numberthis \label{eq:almexplicit}
\end{align*}
where
\begin{equation}
C_{\ell m} \equiv \sqrt{\frac{\ell^2 -m^2}{4\ell^2 -1}}.
\end{equation}
The first term denotes a slight change of the rest-frame multipole with $\ell'=\ell$, and the rest denote the induced multipoles that are leaked from rest-frame multipoles of $\ell'\ne \ell$. One can check that e.g. $\mathcal{K}^{0}_{1}$ and $\mathcal{K}^{0}_{2}$ with $m=0$ are equivalent to the expressions for $\delta T_{1}'(\nu')$ and $\delta T_{2}'(\nu')$ in equation~(\ref{eq:multipoles}), respectively, for an isotropic 21-cm background.

It is further required to rotate the axes, if the motion of the observer against the foreground (e.g. Milky Way) is different from that against the background rest frame (e.g. CMB and the cosmic 21-cm background), and one were to align the $z$-axis along the velocity against the background rest frame. The rotated version of equations (\ref{eq:almtrans}) and (\ref{eq:almexplicit}) will then simply be
\begin{equation}
{a'}_{\ell m}(\nu'|\hat{z}_1) = \sum_{m'=-\ell}^{m'=\ell} D^{\ell *}_{m' m}(\hat{z}_2\rightarrow\hat{z}_1) {a'}_{l m'}(\nu'|\hat{z}_2),
\label{eq:almWigner}
\end{equation}
where $D^{\ell}_{m m'}(\hat{z}_2\rightarrow\hat{z}_1)$ is the Wigner-D matrix that represents the rotation of the axes involving the change of the $z$-axis direction from $\hat{z}_2$ to $\hat{z}_1$, such that
\begin{equation}
Y^{\ell}_{m}(\hat{n}|\hat{z}_1) = \sum_{m'=-\ell}^{m'=\ell} D^{\ell *}_{m m'}(\hat{z}_2\rightarrow\hat{z}_1) Y^{\ell}_{m'}(\hat{n}|\hat{z}_2).
\label{eq:Wigner}
\end{equation} 
Note the interchange of $m'$ and $m$ in equation (\ref{eq:almWigner}) and (\ref{eq:Wigner}), and also the fact that both are the complex conjugated ones. When there are multiple components in the foreground and the background against which the observer velocity ${\bf v}_i$s are in general different, and we take some preferred $z$ direction to define multipoles, we have
\begin{equation}
{a'}_{l m}(\nu'|\hat{z}) = \sum_{i}D^{\ell *}_{m' m}(\hat{\beta}_i\rightarrow\hat{z}) {a'}_{l m'}(\nu'|\hat{\beta}_i) = \sum_{i,\ell'}D^{\ell *}_{m' m}(\hat{\beta}_i\rightarrow\hat{z}) \mathcal{K}^{\ell'}_{\ell}(\beta_i) a_{\ell' m'}(\nu'|\hat{\beta}_i).
\label{eq:almWignerAberr}
\end{equation}
In equations (\ref{eq:almWigner} -- \ref{eq:almWignerAberr}), the condition $|\hat{a}$ means that $\hat{z}=\hat{a}$. In case a radiation field is isotropic and an observer is moving at velocity $\beta_i \hat{\beta}_i$, that radiation component will only pick up $a_{\ell' m'=0}$ terms and the rest with $m'\ne 0$ will be zero. We practically use $i\in\{\rm 21cm,EG,MW,CMB\}$, denoting 21-cm, extra-galactic, galactic and CMB components, respectively. 21cm, EG, and CMB components are all assumed isotropic and thus share the common $z$-axis; the MW component is the only one to apply the Wigner-D matrix to in equation (\ref{eq:almWignerAberr}).
We take the solar velocity against the MW, $V_{\odot,\rm MW}$=($V_R$, $V_\phi$, $V_Z$)=($14.1\pm 1.7$, $248.5\pm 0.5$, $8.49\pm 0.27$) km s$^{-1}$ that was compiled by combining results of \citet{Reid2020} and \citet{Gravity2019}, as in \citet{Malhan2020}. This translates to, in the galactic coordinate, the direction of $\beta_{\odot,\rm MW}$ given by ($l$, $b$)=($86.75\pm 0.39^\circ$, $1.954 \pm 0.062^\circ$). The numerical values of the Wigner-D matrices are shown in the Appendix.

We can roughly estimate how much the radiation-rest multipoles of the foreground contaminates the observed 21-cm signal. Roughly, in the frequency range of interest, $\left| \delta T_b \right| \sim 10^{-4} T^{({\rm FG})}$. The low-$\ell$ multipoles of $T^{({\rm FG})}$, especially those of the Milky Way, are more or less similar in amplitude. Therefore, even a slight leakage of some multipoles may dominate over $\left| \delta T_b \right|$. While equation (\ref{eq:almexplicit}) is truncated at $\mathcal{O}(\beta^2)$, the general trend is that $a_{(\ell + \Delta \ell)m}(\nu')$ is leaked into ${a'}_{\ell m}(\nu')$ at $\mathcal{O}(\beta^{\left|\Delta \ell\right|})$. As $\beta \sim 10^{-3}$ against the center of the Milky Way and also against the CMB rest frame, $\left| \delta {T'}_1 \right|$ and $\left| \delta {T'}_2 \right|$ are of $\mathcal{O}(10^{-7}T^{({\rm FG})})$ and $\mathcal{O}(10^{-10}T^{({\rm FG})})$, respectively. Therefore, $\left| \delta {T'}_1 \right|$ will be subdominant to the induced dipole from the intrinsic Milky Way multipoles in $\ell'$=\{0, 2, 3\}, and similarly $\left| \delta {T'}_2 \right|$ will be so with $\ell'$=\{0, 1, 3, 4, 5\}. In order to calculate these accurately, we need to go to higher order in $\beta$ than equation (\ref{eq:almexplicit}) and implement the methods used in \citet{Chluba2011}, and also \citet{Jeong2014} when masking the sky is required.\footnote{{Note, however, that} we do not take into account masking in this study, which in practice is an important source of complication.}

Fortunately, the spectral smoothness of foregrounds is also inherited in all induced multipoles. Therefore, we may not need to quantify all the induced multipoles of each foreground component with strong intrinsic multipoles. The foreground removal procedure, in practice, will naturally remove the induced multipoles as well. If any low-order intrinsic multipole of the foreground had a peculiar spectral structure, extracting the cosmic 21-cm dipole and quadrupole would be extremely difficult. For the Fisher analysis in this paper, once a spectral smoothness of the foreground is assumed, it is sufficient to just take the expansion of ${a'}_{\ell m}(\nu')$ in equation (\ref{eq:almWignerAberr}) up to $\mathcal{\beta}$, or equivalently considering the terms only of $a_{\ell' m}$ with $\ell'-\ell=\{-2,-1,\,0,\,1,2\}$. In such a limited case, we may use the approximation
\begin{equation}
{a'}_{\ell m}(\nu'|\hat{\beta}_i)
\simeq a_{\ell m}(\nu'|\hat{\beta}_i) 
+\mathcal{K}^{l+2}_{l} a_{(\ell+1) m}(\nu'|\hat{\beta}_i)
+\mathcal{K}^{l+1}_{l} a_{(\ell+1) m}(\nu'|\hat{\beta}_i)
+\mathcal{K}^{l-1}_{l} a_{(\ell-1) m}(\nu'|\hat{\beta}_i)
+\mathcal{K}^{l-2}_{l} a_{(\ell-1) m}(\nu'|\hat{\beta}_i)\,,
\label{eq:simpleleak}
\end{equation}
for the Fisher analysis but still study the impact of $\beta$.

\subsection{Observed signatures}\label{sec:observed}

Following previous subsections, the observed monopole, dipole and quadrupole spherical harmonic coefficients take the form
\be
    a_{00}^{\rm obs}(\nu',\hat{z})&\simeq&\delta T_0'(\nu') + \aeg_{00}(\nu'|\hat{z})\!+\! D_{00}^{0\,*}\,\big[\amw_{00}(\nu'|\hat{\beta}_{\rm MW})+\sum\limits_{\ell'=1}^{2} \mathcal{K}_{\ell}^{\ell+\ell'}\big|_{\ell=0}\amw_{\ell'0}(\nu'|\hat{\beta}_{\rm MW})\big]
    \!+\!\sqrt{4\pi}\left[T_{\rm noise}(\nu')\!+\!\Delta T_{\rm CMB}\right],\,\,\,\nonumber\\
    a_{10}^{\rm obs}(\nu',\hat{z})&\simeq&\delta T_1'(\nu')+\mathcal{K}_\ell^{(\ell-1)}\big|_{\ell=1}\aeg_{00}(\nu'|\hat{z})+\sum\limits_{m=-1}^{-1} D_{m0}^{1*}A_m(\nu'|\hat{\beta}_{\rm MW})+\sqrt{4\pi}T_{\rm noise}(\nu')+\sqrt{\frac{4\pi}{3}}\beta\Delta T_{\rm CMB}\,,\nonumber\\
    a_{1m}^{\rm obs}(\nu',\hat{z})&\simeq&\sum\limits_{m'=-1}^{1} D_{m'm}^{1*}A_{m'}(\nu'|\hat{\beta}_{\rm MW})+\sqrt{4\pi}T_{\rm noise}(\nu')\,,\nonumber\\
    a_{20}^{\rm obs}(\nu',\hat{z})&\simeq&\delta T_2'(\nu')+\mathcal{K}_\ell^{(\ell-2)}\big|_{\ell=2}\aeg_{00}(\nu'|\hat{z})+\sum\limits_{m=-2}^{2} D_{m0}^{2*}B_m(\nu'|\hat{\beta}_{\rm MW})+\sqrt{4\pi}T_{\rm noise}(\nu')+\sqrt{\frac{16\pi}{45}}\beta^2\Delta T_{\rm CMB}\,,\nonumber\\
    a_{2m}^{\rm obs}(\nu',\hat{z})&\simeq&\sum\limits_{m'=-2}^{2} D_{m'm}^{2*}B_{m'}(\nu'|\hat{\beta}_{\rm MW})+\sqrt{4\pi}T_{\rm noise}(\nu')\,,
\ee
where 
\be
    A_{-1}&=&\amw_{1-1}+\mathcal{K}_{\ell}^{(\ell+1)}\big|_{\ell=1}\amw_{2-1}+\mathcal{K}_{\ell}^{(\ell+2)}\big|_{\ell=1}\amw_{3-1}\,,\nonumber\\
    A_0&=&\amw_{10} + \mathcal{K}_{\ell}^{(\ell+1)} \big|_{\ell=1}\amw_{20} +\mathcal{K}_{\ell}^{(\ell-1)}\big|_{\ell=1}\amw_{00}+ \mathcal{K}_{\ell}^{(\ell+2)} \big|_{\ell=1}\amw_{30}\,,\nonumber\\
    A_{1}&=&\amw_{11}+\mathcal{K}_{\ell}^{(\ell+1)} \big|_{\ell=1}\amw_{21}+\mathcal{K}_{\ell}^{(\ell+2)}\big|_{\ell=1}\amw_{31}\,,
\ee
and
\be
    B_{-2}&=&\amw_{2-2}+ \mathcal{K}^{(\ell+1)}_2\big|_{\ell=2} \amw_{3-2}+ \mathcal{K}^{(\ell+2)}_2\big|_{\ell=2} \amw_{4-2}\,,\nonumber\\
    B_{-1}&=&\amw_{2-1} + \mathcal{K}_{\ell}^{(\ell+1)} \big|_{\ell=2}\amw_{3-1}+ \mathcal{K}_{\ell}^{(\ell-1)} \big|_{\ell=2}\amw_{1-1}+ \mathcal{K}_{\ell}^{(\ell+2)} \big|_{\ell=2}\amw_{4-1}\,,\nonumber\\
    B_{0}&=&\amw_{20} + \mathcal{K}_{\ell}^{(\ell+1)} \big|_{\ell=2}\amw_{30} + \mathcal{K}_{\ell}^{(\ell-1)} \big|_{\ell=2}\amw_{10}+ \mathcal{K}_{\ell}^{(\ell+2)} \big|_{\ell=2}\amw_{40}+ \mathcal{K}_{\ell}^{(\ell-2)} \big|_{\ell=2}\amw_{00}\,,\nonumber\\
    B_{1}&=&\amw_{21} + \mathcal{K}_{\ell}^{(\ell+1)} \big|_{\ell=2}\amw_{31} + \mathcal{K}_{\ell}^{(\ell-1)} \big|_{\ell=2}\amw_{21} + \mathcal{K}_{\ell}^{(\ell+2)} \big|_{\ell=2}\amw_{41}\,,\nonumber\\
    B_{2}&=&\amw_{22} + \mathcal{K}_{\ell}^{(\ell+1)} \big|_{\ell=2}\amw_{32} + \mathcal{K}_{\ell}^{(\ell+2)} \big|_{\ell=2}\amw_{42}\,, 
\ee
where we omitted showing the dependency of the kernels on the velocity for brevity, and $\hat{\beta}_{\rm MW}\equiv\hat{\beta}_{\odot,\rm MW}$. 

We model the global {galactic and extra-galactic} foreground signals with a 4th order polynomial\footnote{{The fourth order polynomial was chosen to capture the smooth 21-cm foregrounds sufficiently accurately. Observing the smoothness of the foreground profile, we have found the 4th order poynomial provides a sufficient and excellent fit to foreground spectrum of spherical multipoles for our purposes in this paper.}} satisfying 
\begin{equation}\label{eq:foreground}
    a_{\ell m}^{(J)}(\nu')=\sum\limits_{i=0}^{4}a_{\ell m,i}^{(J)}\left[\ln(\nu'/\nu_{0})\right]^i\,,
\end{equation}
where $J\in\{\rm EG,MW\}$ spanning over extra-galactic and galactic spherical harmonic coefficients, respectively. We calculate the fiducial values for the coefficients parameterizing the Milk-Way foreground using the PyGDSM\footnote{The PyGDSM code can be accessed at \hyperlink{https://github.com/telegraphic/pygdsm}{github.com/telegraphic/pygdsm}. We take the most recent, Low Frequency Sky Survey (LFSS) data measured by the Low Frequency Array \citep{Dowell2017}. PyGDSM allows the full-sky, full-frequency 3D data that is extrapolated from the LFSS data by the principal component analysis.} code~\citep{2008MNRAS.388..247D,Dowell2017,2017MNRAS.464.3486Z} as we detail in Table~\ref{tab:coefficients} which match the anticipated total foreground in ~\citet{deOliveira-Costa:2008cxd}, \citet{Rogers:2008vh}, \citet{2009A&A...500..965B}, \citet{Bernardi:2016pva}, and \citet{Dowell2017}. We consider only the monopole of the extra-galactic foreground assuming higher multipoles of this signal will be smaller by many orders of magnitude due to the observed homogeneity and isotropy of the large-scale structure. The foreground monopole, dipole and quadrupole spectra we consider are shown on the right panels of Fig.~\ref{fig:FIG1} with  solid curves. The dashed lines on these panels correspond to the kinematic contributions to signatures given in Section~\ref{sec:observed}.

\subsection{The 21-cm global signal\label{sec:21cmfast}}
\begin{figure}[t!]
\centering
\includegraphics[width=0.9\linewidth]{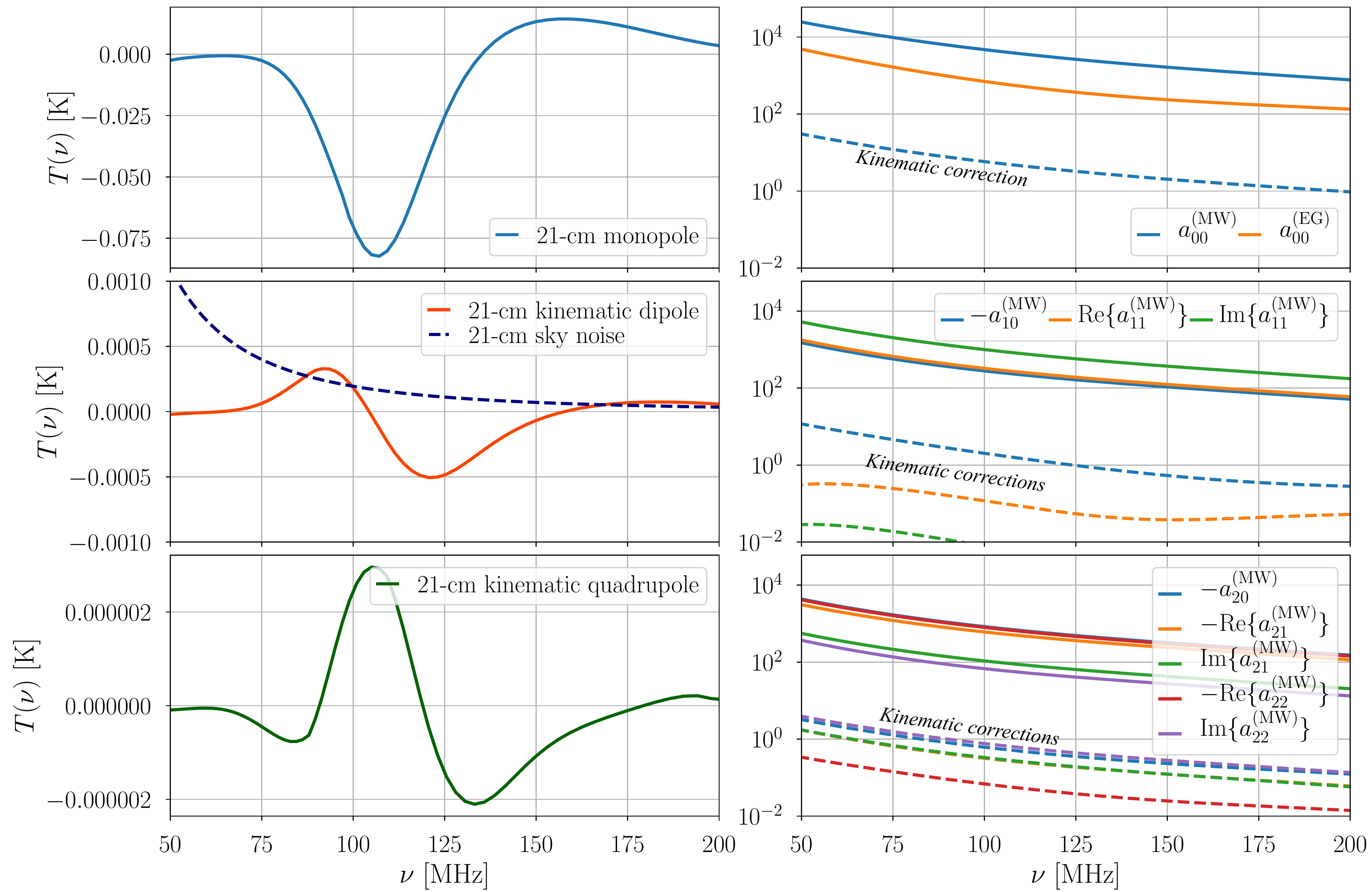}
\caption{(\textit{Left panels}) From top to bottom, the monopole, dipole and quadrupole contributions to the global 21-cm signal as a function frequency. The monopole (solid blue curve) is calculated using 21cmFAST with astrophysical parameter \{$f_{*,10}$, $\alpha_*$, $\alpha_{\rm esc}$, $L_x$, $t_*$, $\,M_{\rm turn}$\} set equal to \{0.05, 0.5, -0.5, 40, 0.5, 8.7\} (default choices in 21cmFAST) and assuming \textit{Planck}~2018 cosmology~\citep{Planck:2018nkj}. The dipole and quadrupole signals are calculated using equation \eqref{eq:multipoles}. The purple dashed line on the middle left panel corresponds to $T_{\rm noise}(\nu_0)=0.44$mK statistical measurement noise due to finite integration time at $\nu_0=76$MHz, defined in {equation~(\ref{eq:Tnoise})} with $T_{\rm sky}(\nu_0)=1570$K, $B=0.4$MHz and $t_{\rm int}=1$ year, representative of the EDGES experiment. {Note, however, that the EDGES experiment is dominated by {a} systematic error level of the order of $\gtrsim {20 { \rm mK}}$.} (\textit{Right panels}) The solid lines correspond to galactic and extra-galactic foreground spectra considered in this analysis. Similarly from top to bottom, panels correspond to the monopole, dipole and the quadrupole signals, respectively. The dashed lines with matching colours correspond to the kinematic contributions to the signals, given in Sec.~\ref{sec:observed}. The super-scripted MW and EG labels on the coefficient spectra correspond to Milky-Way and extra-galactic foregrounds, respectively. ``Re" and ``Im" represent the real and the imaginary parts of the given multipole amplitude. The kinematic corrections shown in middle and lower right panels also include the relativistic correction from the extra-galactic signal.}
\label{fig:FIG1}
\end{figure}

The left panels in Fig.~\ref{fig:FIG1} correspond to the cosmological 21-cm global signal. We model the astrophysical parameters following the 21cmFAST code~\citep{Mesinger:2010ne,Mesinger:2007pd} and include: $f_{*,10}$, the fraction of galactic gas in stars for $10^{10}$ solar mass halos; $\alpha_*$, the power-law index of fraction of galactic gas in stars as a function of halo mass; $a_{\rm esc}$, the power-law index of escape fraction as a function of halo mass; $L_x$, the specific X-ray luminosity per unit star formation escaping host galaxies; and $t_*$, the fractional characteristic time-scale (fraction of Hubble time) defining the star-formation rate of galaxies following the definitions in Ref.~\citep{Park:2018ljd}. The  monopole spectra are shown on the top left panel of Fig~\ref{fig:FIG1} with the solid blue curve. The orange and dark-green solid curves on the panels below correspond to dipole and quadrupole contributions to the observed cosmological 21-cm global signal. The dashed purple line corresponds to a measurement noise anticipated to match EDGES sensitivity at 1 year of observation time modulo systematic
uncertainties.

\section{Forecasts\label{sec:Forecasts}}

We assume the monopole, dipole and quadrupole moments can be measured by weighting the observed brightness temperature maps with spherical harmonics $Y_{00}(\hat{n}), Y_{1m}(\hat{n})$ and $Y_{2m}(\hat{n})$, respectively. Throughout we assume measurements have access to the full-sky and assume no errors on the \textit{direction} of Earth's velocity with respect to CMB and the Milky Way\footnote{In practice, the incomplete sky coverage will introduce different non-unity coefficients in-front of the observed monopole, dipole and quadrupole components; in a way depending on the direction and the measurement error of the velocities. We will incorporate such effects in an upcoming study.}. In order to assess the information content of the dipole and the quadrupole signals we define the standard information matrix for a single multipole experiment as 
\begin{equation}\label{eq:fisher}
    F_{ij}=\sum\limits_{X,n=1}^{N_{\rm channel}} \left[\frac{\partial T_X^{\rm obs}(\nu_n)}{\partial \alpha_i}\frac{\partial T_X^{\rm obs}(\nu_n)}{\partial \alpha_j}\right]\frac{1}{T_{\rm noise}^2 (\nu_n)}\,,
\end{equation}
where the parameter set $\{\alpha_i\}$ includes both foreground and signal model parameters, $X\in\{M,D,Q\}$ spans over the observed monopole, dipole and quadrupole moments, $T_{\rm noise}$ is the thermal noise given by the radiometer equation
\begin{equation}
    T_{\rm noise}^2 (\nu_n)=\frac{T_{\rm sky}^2(\nu_n)}{B t_{\rm int}}\,,
\label{eq:Tnoise}
\end{equation}
and $B$ is the bandwidth in frequency running between $[\nu_{\rm min},\nu_{\rm max}]$ in a given frequency bin centered at the frequency $\nu_n$, $N_{\rm channel}$ is the number of frequency bins the signal is divided in, $t_{\rm int}$ is the integration time and $T_{\rm sky}(\nu)=a_0(\nu/\nu_0)^{-2.5}$ where we set $a_0=1570$K and $\nu_0=76$MHz. {The `obs' superscript indicates that we use the temperature multipole moments of the net observed signal in constructing the Fisher matrix. Each $T_X^{\rm obs}$ for $X\in\{M,D,Q\}$ is dominated by the corresponding multipole moment of the Milky Way signal.}

\begin{figure}[t!]
\centering
\includegraphics[width=0.47\linewidth]{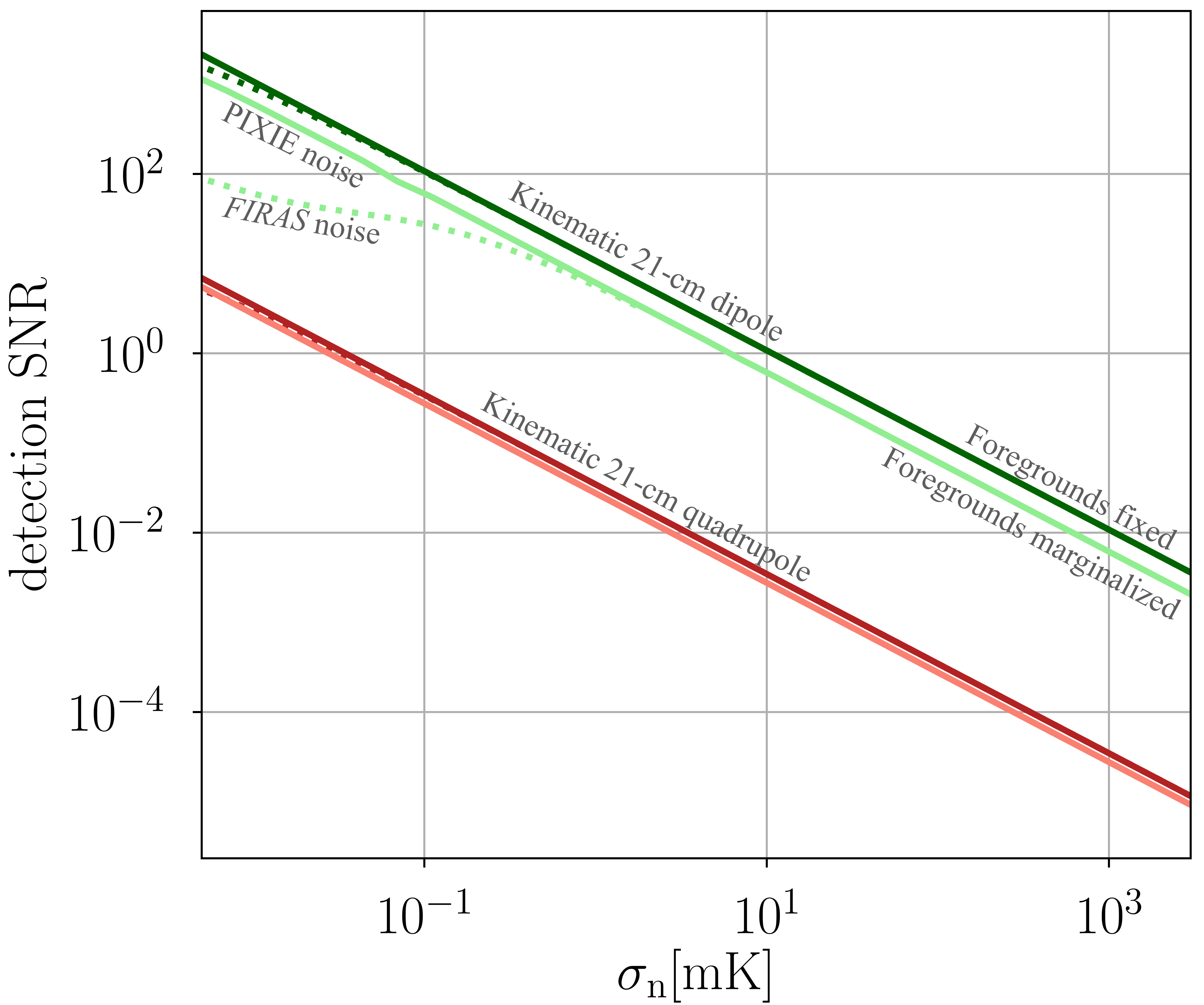}
\caption{The signal-to-noise (SNR) of the dipole and quadrupole contributions to the 21-cm global signal. The dotted lines corresponds to assuming the FIRAS constraints on the mean CMB temperature satisfying $\sigma(T_{\rm CMB,0})=0.57$mK. The solid lines corresponds to forecasted accuracy of the proposed PIXIE experiment $\sigma(T_{\rm CMB,0})=1$nK. The $x$-axis corresponds to RMS noise of the 21-cm temperature satisfying $\sigma_n=T_{\rm sky}(\nu_0)/\sqrt{B t_{\rm int}}$. The RMS noise satisfies $\sigma_n(\nu_0)=0.44$mK for $T_{\rm sky}(\nu_0)=1570$K, bandwidth value of $B=0.4$MHz as reported by the EDGES {data release}, and $t_{\rm int}=1$ year, representative of the EDGES data. {The upper lines correspond to the SNR of the kinematic 21-cm dipole signal. Here, the darker green lines correspond to fixing the foreground spectra while lighter green lines correspond to marginalizing over coefficients that parameterize the Milky-Way and extra-galactic foregrounds as well as the astrophysical parameters as described in the text. We find that precise measurement of the CMB temperature (e.g.~by~PIXIE) may improve the detection prospects of the kinematic 21-cm dipole if foreground model parameters {are} marginalized. The lower lines correspond to the SNR of the kinematic 21-cm quadrupole. Here we find the SNR does not depend strongly on whether the foreground and astrophysical parameters are marginalized or fixed. For both the dipole and quadrupole, we have found that the detection SNR do not depend strongly on the choices of priors on the foreground parameters we consider in this paper.}}
\label{fig:FIG2}
\end{figure}

\begin{figure}[t!]
\centering
\includegraphics[width=0.7\linewidth]{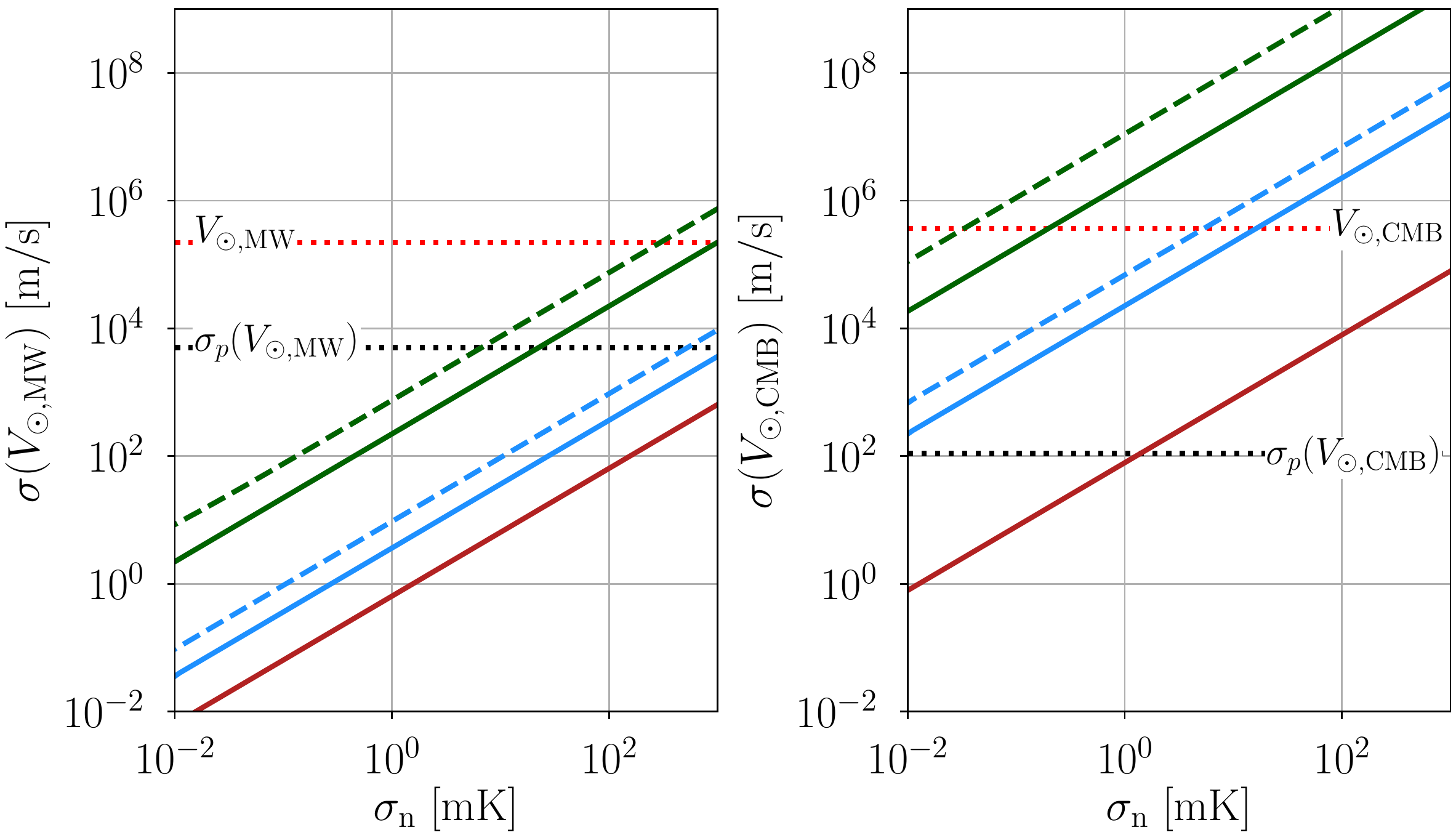}
\caption{Forecasts of the measurement accuracy on the velocity from the dipole and the quadrupole signatures. The dashed (solid) lines corresponds to assuming the FIRAS (PIXIE) constraints on the mean CMB temperature as discussed in the text. The left panel shows the constraints on Earth's velocity with respect to the Milky Way $V_{\odot\rm MW}\simeq250\,{\rm km/s}$ with a conservative measurement error  $\sigma_p(V_{\odot,\rm MW})\simeq5\,{\rm km/s}$ coming from \citet{Reid2014}. The right panel corresponds to measurement accuracies on Earth's velocity with respect to the extra-galactic background, which we take to be the velocity with respect to the CMB surface $V_{\odot,\rm CMB}\simeq369.89{\rm km/s}$. The black dotted line on this panel corresponds to the current strongest measurement accuracy from \textit{Planck} at $\sigma_p(V_{\odot,\rm CMB})\simeq0.11{\rm km/s}$~\citep{Planck:2018nkj}. In both panels the green and blue lines correspond to considering up to the dipole and dipole plus quadrupole signals, respectively. Here, we fix the parameters defining the spectrum of harmonic coefficients $a_{\ell m,i\neq0}$ as defined in Eq.~\eqref{eq:foreground}, and set $a_{\ell m,i=0}$ with $\ell>0$ to be free parameters with 10 percent priors (this or `B1' choice of priors as described in the text). The dashed green and blue lines corresponding to the errors obtained by setting all coefficients to be free parameters, while choosing $10$ percent priors on coefficients satisfying $a_{\ell m,i=0}$ with $\ell>0$ and $1$ percent priors on coefficients satisfying $a_{\ell m,i>0}$ with $\ell>0$ (this is our prior choice `B2'). In both bases, the parameters modelling the spectrum of the Milky-Way and extra-galactic monopole, $a^{\rm MW}_{00,i}$ and $a^{\rm EG}_{00,i}$, are set to be free parameters, for which we assume no priors. We find measurement accuracy of Earth's velocity with respect to Milky Way (CMB) assuming B2 choice of priors reduce by a factor of $\sim2$ ($\sim10$). The red lines correspond to fixing all coefficients other than $a_{00,i}$. We find measurement of the dipole and quadrupole signals with ongoing and upcoming radio experiments can potentially improve the measurement accuracy on Earth's velocity with respect to the Milky Way while improving the velocity with respect to CMB using these observables may prove difficult unless the spectral profiles of the harmonic coefficients are well known from external measurements or theory. The $x$-axis corresponds to RMS noise of the 21-cm temperature satisfying $\sigma_n=T_{\rm sky}(\nu_0)/\sqrt{B t_{\rm int}}$. The RMS noise satisfies $\sigma_n=0.44$mK for $T_{\rm sky}(\nu_0)=1570$K, $B=0.4$MHz and $t_{\rm int}=1$ year, representative of the EDGES data.}
\label{fig:FIG3}
\end{figure}

We take the velocity of the solar system in the background rest frame to be $V_{\odot,\rm CMB} \simeq369.82\,{\rm km/s}$ as measured by \textit{Planck}~\citep{Planck:2018nkj}. We note that even though the amplitude and direction of $V_{\odot,\rm CMB}$ are well constrained by the CMB temperature anisotropy, many other probes such as the quasar observables result in much larger values and somewhat different directions \citep{Secrest2021}. Here we limit ourselves to the CMB-measured local velocity for comparison with the measurement prospects from the global 21-cm signal, where the error on the former is $\sim0.11\,{\rm km/s}$~\citep{Planck:2018nkj}. In what follows we assume both the foregrounds and 21-cm signal experience the same Doppler and aberration effects due to this velocity and consider errors, $0.57$mK and $1$nK, on the mean CMB temperature $T_{\rm CMB,0}$ measured by FIRAS~\citep{Fixsen:1996nj} and anticipated by PIXIE~\citep{2011JCAP...07..025K}, respectively.

Fig.~\ref{fig:FIG2} shows the signal-to-noise ratio (SNR) of the dipole and the quadrupole components of the 21-cm signal. Here, we parameterize the dipole and quadrupole contributions to the observed intensity as $A_{1}\delta T_{1}$ and $A_{2}\delta T_{2}$, respectively, and forecast the measurement accuracy on amplitudes $A_1$ and $A_2$, whose fiducial values are unity. The solid (dotted) lines correspond to considering the CMB monopole measured by PIXIE (FIRAS). In the case the foreground and astrophysical parameters are fixed, we find the SNR does not depend on the noise on the CMB monopole significantly. Once the foreground and astrophysical parameters are marginalized, we find in particular the detection SNR of the kinematic 21-cm dipole becomes sensitive to the CMB noise. Overall, we find a radio experiment measuring the 21-cm dipole at the sensitivity of EDGES could potentially detect the cosmological kinematic dipole signal while detecting the cosmological 21-cm quadrupole requires over around an order of magnitude reduction in the statistical measurement noise, and that for low-noise radio measurements, the detection SNR of the kinematic 21-cm dipole may improve if the measurement precision of the CMB monopole improves.\footnote{Note that throughout we consider only the statistical noise, which depends on the integration time. Present challenges however also include properly accounting for systematic errors, whose modelling will likely require future work in order to achieve the measurement of the relevant quantities in this paper.} 

Fig.~\ref{fig:FIG3} shows our forecasts on the measurement accuracy of the velocity from 21-cm measurements. The constraints on $V_{\odot, J}$ where $J=\{\rm MW, EG\}$ are significantly affected by the uncertainty on the foreground parameters modelled in Eq.~\eqref{eq:foreground}. The blue and green lines on the right panel (left panel) demonstrates the error on Earth's velocity with respect to the 21-cm background (Milky Way) as a function of 21-cm measurement noise, once the foreground parameters are marginalized using a Fisher matrix as given in Eq.~\eqref{eq:fisher}. The solid green and blue lines in both panels correspond to our `baseline~1' (B1) prior choice where we assume 10 percent priors on coefficients $a_{\ell m,i}$ satisfying $i=0$ and fix coefficients satisfying $i\neq0$ and $\ell\neq0$ to their fitted values from PyGDSM simulation~\citep{2016ascl.soft03013P} of the galactic foreground. We set coefficients $a_{00,i}$ free without assuming any priors. The blue lines correspond to including both dipole and quadrupole signals in the analysis, while the green lines correspond to including only the dipole signal. The red lines on the right panel correspond to fixing \textit{all} coefficients $a_{\ell m,i}$ satisfying $\ell\neq0$. Fixing these coefficients result in a factor of $\sim2$ improvement for the measurement of Earth's velocity with respect to Milky Way (red solid line on the left panel) compared to our B1 prior choice including the quadrupole signal. The dashed green and blue lines correspond to our `baseline~2' (B2) prior choice, where we set all parameters to be free with 100 percent priors on the amplitude of each spherical harmonic coefficient $a_{\ell m,0}$ and set 10 percent priors on coefficients satisfying $a_{\ell m,i}$ with $i>0$. 

The B2 prior choice leads to a factor of $\sim2-3$ reduction in the measurement accuracy of Earth's velocity with respect to Milky Way and CMB compared to B1. The red dotted horizontal line on the right panel corresponds to $369.86{\rm km/s}$, suggesting high-significance measurement of the velocity could still be achieved with ongoing and upcoming 21-cm experiments. The dotted black horizontal line on the right panel corresponds to the error on the velocity provided by the local dipole measurement by \textit{Planck}~\citep{Planck:2018nkj}. The red dotted line on the left panel corresponds to the measured velocity of Earth with respect to the Milky-Way $V_{\odot,\rm MW}\simeq250{\rm km/s}$ and the solid dotted line in the same panel corresponds to the error on this measurement $\sim5{\rm km/s}$ \citep{Reid2014}. We find the large-valued higher order multipoles of the galactic foreground make measuring the quadrupole signal valuable for improving the measurement precision of $V_{\odot, \rm MW}$. The choices of priors affect significantly the prospects of detecting velocities with the radio signal due to degeneracies between velocities and foreground coefficients. Lastly, the left panel assumes \textit{Planck} prior on the cosmological velocity $\sigma_p(V_{\odot, \rm CMB})=0.11{\rm km/s}$ while setting $V_{\odot, \rm MW}$ to be a free parameter without a prior. The right panel assumes the prior $\sigma_p(V_{\odot, \rm MW})=5{\rm km/s}$ while setting $V_{\odot, \rm CMB}$ free without a prior. 

\begin{figure}[t!]
\centering
\includegraphics[width=0.85\linewidth]{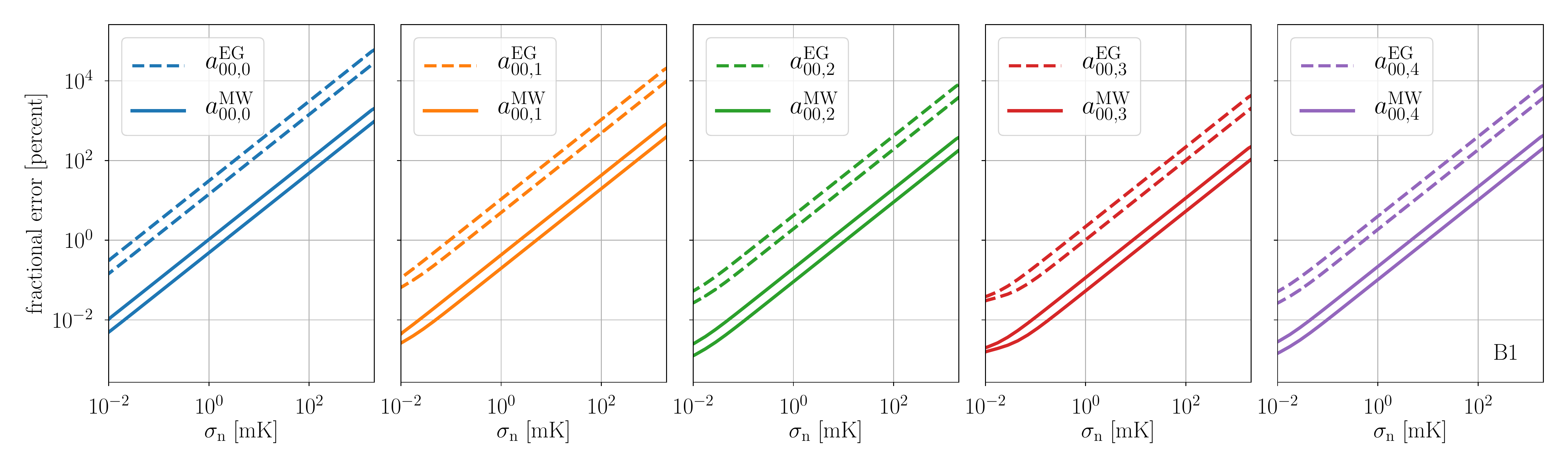}\vspace*{-0.25cm}
\caption{The figure shows the fractional error (in percentage: the $y-$axis shows $100\times\sigma(p_i)/p_i$) on the parameters that characterize the spectral profile of the galactic (extra-galactic) monopole with solid (dashed) curves as a function of 21-cm noise. The upper (lower) lines for both line styles correspond to including the monopole and dipole signals (monopole, dipole and quadrupole signals). For these panels, we consider the prior choice we defined as `B1' in the text, where we fix all parameters except those that parameterize the monopoles and the amplitudes of the higher-order spherical harmonic coefficients. We find for this choice, addition of the quadrupole with respect to adding only the dipole improves the measurement quality of the monopole by around a factor of 2.}
\label{fig:FIG4}\vspace*{-0.4cm}
\end{figure}

\begin{figure}[t!]
\centering
\includegraphics[width=0.85\linewidth]{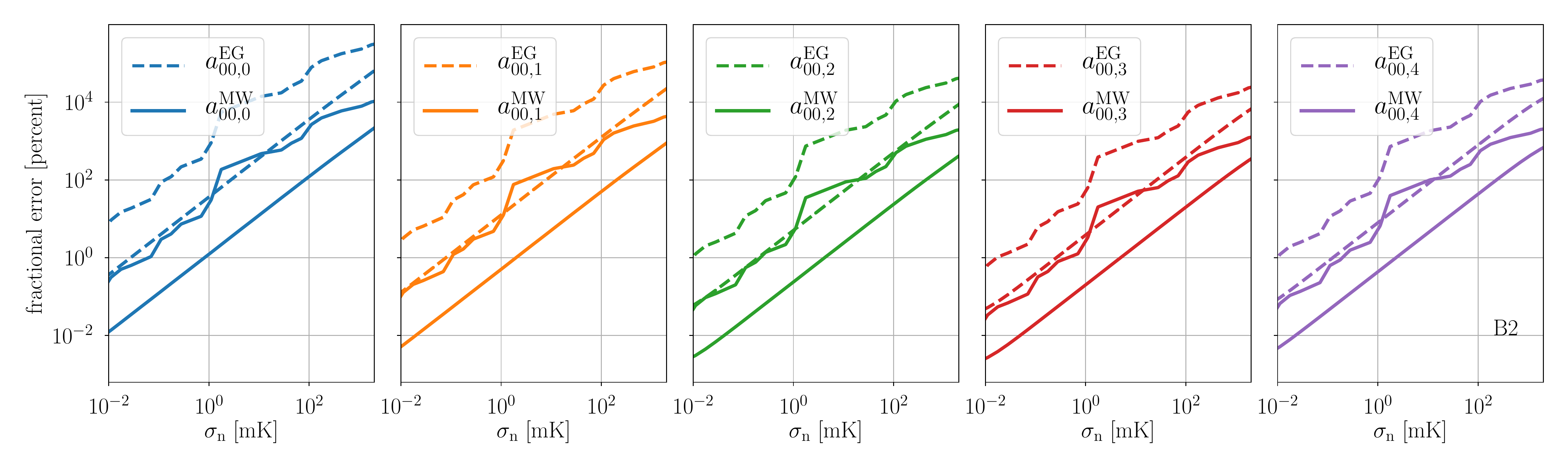}\vspace*{-0.25cm}
\caption{Figure similar to Fig.~\ref{fig:FIG4} with the exception that here we consider the prior choice we labelled as `B2' in the text, where we set all parameters free with $100$ ($10$) percent priors on the amplitudes (all else). We find for these less restrictive priors, the quadrupole improves the measurement accuracy of the galactic and extra-galactic monopole around a factor of $5-10$, in this example.}
\label{fig:FIG5}\vspace*{-0.4cm}
\end{figure}

\begin{figure}[t!]
\centering
\includegraphics[width=1\linewidth]{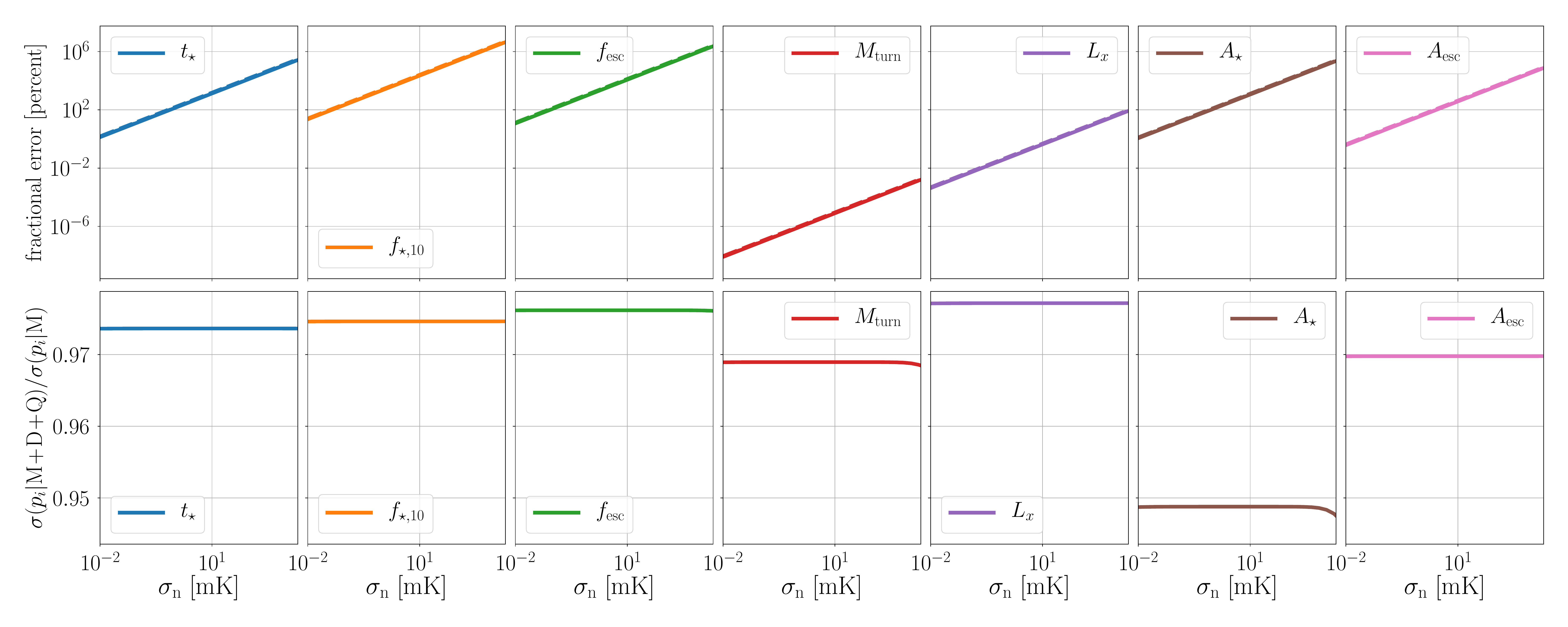}
\caption{The panels show the relative percent improvement on astrophysical parameters determining the 21-cm signal and the Milky-Way background monopole for $\sigma_{\rm n}=3{\rm mK}$. The left panel corresponds to the percent improvement in errors from adding the dipole and quadrupole signals to the measurement of the monopole. Here, the improvement is sourced dominantly by the dipole signal. The middle panel corresponds to fractional improvement (in percentage) from adding the quadrupole signal to the measurement of monopole and dipole signals for the prior choices B1 as described in the text. Here, we find the improvements from adding the quadrupole signal is marginal (around ~85 percent for all parameters of the polynomial expansion of the monopole spectra). The right panel corresponds to the fractional improvement (in percentage) from adding the quadrupole to the measurement of monopole and dipole signals for the B2 prior choice we describe in the text. Without strong priors on the parameters modelling the 21-cm background, the measurement of the quadrupole can be shown to lead to a factor of $\sim5-9$ improvement on the measurement precision of the Milky-Way foreground (around $\sim500-900$ percent improvement).}
\label{fig:FIG6}\vspace*{-0.4cm}
\end{figure}

\begin{figure}[t!]
\centering
\includegraphics[width=1\linewidth]{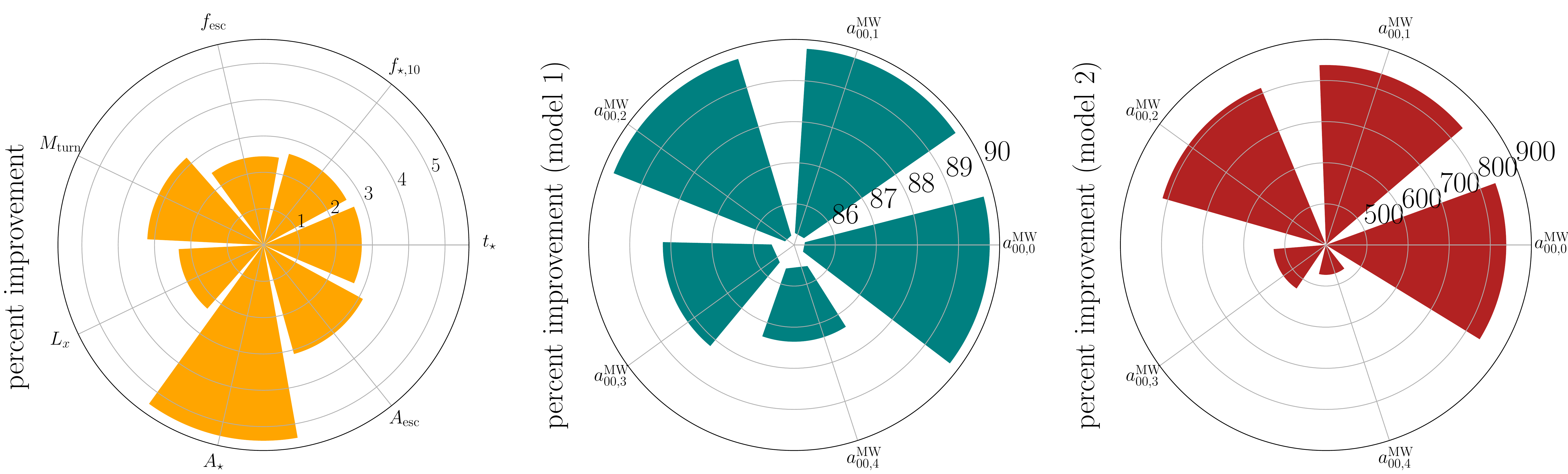}\vspace*{-0.25cm}
\caption{The top panels show the fractional errors on the 21-cm astrophysical parameters as a function of 21-cm measurement noise, otherwise similar to Fig.~\ref{fig:FIG4}. Bottom panels correspond to the error ratio between measuring all three monopole, dipole and quadrupole signals and measuring only the monopole. We find that adding the quadrupole to the monopole and dipole signals improves measurement precision on the 21-cm astrophysical parameters marginally, by a few per cent (see Fig.~\ref{fig:FIG6}).}
\label{fig:FIG7}
\end{figure}

Another notable scientific gain from measuring the dipole and quadrupole contributions to the radio foreground is the ability to distinguish the galactic contribution from the extra-galactic foreground signal. Throughout we have assumed vanishing contribution to the extra-galactic foreground from higher multipoles setting $T^{\rm (EG)}(\nu,\hat{n})\simeq a_{00}^{\rm (EG)}Y_0^{0}(\hat{n})$ which is degenerate with the galactic foreground if the kinematic contributions are omitted and only the monopole is measured. By probing the dipole and the quadrupole signals, however, one can potentially separate the extra-galactic foreground from the Milky Way through measuring the leakage of the extra-galactic monopole into dipole and quadrupole signals, due to the dependence of the extra-galactic and Milky-Way foregrounds on different velocities $V_{\odot,\rm CMB}$ and $V_{\odot,\rm MW}$, respectively. Using an experiment with sensitivity for these signals, one can in principle model the extra-galactic monopole and the galactic foregrounds jointly and measure their spectra. Using our B1 prior choices defined above, we find measuring the dipole and quadrupole signals allow separate measurements of the extra-galactic and Milky-Way monopole spectra parameters despite the degeneracy of the parameters modelling the different spectra. We show the fractional errors on the extra-galactic and Milky-Way monopole spectra parameters in Figures \ref{fig:FIG4} and \ref{fig:FIG5}.

Adding the measurement of the quadrupole signal increases the measurement accuracy of the astrophysical background significantly (around 85 percent for spherical harmonic coefficients $a^{\rm MW}_{00,i}$ for $i\in[0-4]$, for example, and equivalently for $a^{\rm EG}_{00,i}$) as shown in the middle panel of Fig.~\ref{fig:FIG6}. Setting all coefficients to be free parameters and taking our B2 prior choices, we find the relative improvement from adding the quadrupole signal also increases, leading to a $\sim500-900$ percent (around a factor of $\sim5-9$) improvement on the measurement of the foreground monopole parameters compared to adding the dipole signal only, as shown in the right panel in Fig.~\ref{fig:FIG6}. We also find the measurement accuracy of the astrophysical parameters \{$f_{*,10}$, $\alpha_*$, $\alpha_{\rm esc}$, $L_x$, $t_*$, $\,M_{\rm turn}$\} get $\sim2-5$ percent improvements from considering the kinematic contributions of the 21-cm signal to the higher-order multipoles, in a way less sensitive to the choices of the priors. This is shown in the left panel of Fig.~\ref{fig:FIG6}.  Our parameter errors are in agreement with similar forecasts in literature \citep[see e.g.][]{Park:2018ljd,Mason:2022obt}.   
 Fractional errors on the 21-cm parameters are shown in Fig.~\ref{fig:FIG7}.We find the improvement of the measurement accuracy of these parameters is dominated by the dipole signal. 

For the 21-cm background parameters, however, we stress that the additive information on higher-order multipoles can constrain the 21-cm monopole spectrum significantly by providing a consistency check through equation (\ref{eq:multipoles}) in addition to the estimated measurement accuracy which shows around a factor 2 or greater improvement. The consistency check can be made as follows. One takes the $\delta T'(\nu')$ estimated from a given monopole measurement, and compares the correspondingly estimated dipole and quadrupole spectra to the data measured by some future, dedicated experiment. There are already two existing 21-cm monopole data {sets} from the EDGES and the SARAS-3 experiments, which are mutually inconsistent. At the moment, the SARAS-3 data seems to be more free from systematic errors{, such as the beam chromaticity and the environmental effects,} and consistent with the null {signal} ($\delta T_{b}(\nu')=0$) at $\nu' \le 100\,$MHz, in contrast to the claimed detection of $\sim 500\,$mK absorption trough by the EDGES team~\citep[see~e.g.~][and references therein]{SARAS2022}. The problem with such single-dish experiments is that the derived 21-cm monopole spectrum suffers from the intrinsic ambiguity caused by the uncertainty in the exact form of the foreground spectrum, as demonstrated by~\citet{Hills2018}. After measuring the 21-cm dipole (and better yet the quadrupole as well), one could in principle run consistency tests on these monopole data sets to settle the apparent conflicts, depending on the measurement quality of the former.


\section{Summary and Discussion\label{sec:discussion}}

In this work we have shown that the measurement of radio-frequency dipole and quadrupole signals can potentially improve the precision measurements of Earth's velocity with respect to Milky Way and CMB and allow jointly constraining and distinguishing-between the extra-galactic and the Milky-Way foregrounds. We showed how these improvements depend on our external knowledge on the spectrum of galactic and extra-galactic spherical harmonic multipoles by considering two sets of prior choices on the radio-frequency background spectrum. Furthermore, we demonstrated that the measurements of dipole and quadrupole signals could lead to modest improvements on the 21-cm astrophysics, improving parameter constraints by around few percents. Although we find the improvements on the 21-cm parameters are marginal, the measurement of dipole and quadrupole 21-cm signal can potentially serve as a consistency check of the monopole spectrum.  

There are a few reasonable assumptions that we made. Both CMB and 21-cm background were assumed to be isotropic for low-order multipoles in their shared rest frame. Therefore, both dipole and quadrupole moments of these backgrounds in the observer frame should be  kinematically induced ones, not the intrinsic ones. We also assumed that the full sky could be observed, and accordingly considered the full-sky 3D radio map that dominates every multipole moments in the frequency range of our interest. We considered variation in the solar velocities against the Milky Way and the background radiation, but not the directional variation. 


There may arise some difficulties in actual observation and analysis of these low-$\ell$ multipole moments. Partial sky coverage may be forced from either a ground telescope at one location, or if the spectral contamination by the Milky Way plane and bulge is too noisy. This then could lead to incomplete measurement of the 21-cm multipoles. 
Spectral smoothness of the foreground in each multipole is crucial not only for our proposed observation but also for the high-redshift 21-cm science in general.


What would be the optimal observational strategy?
With high-sensitivity apparatus such as several SKA precursors and the SKA-LOW, one could in principle observe low-order multipoles but again with significantly limited sky coverage from each one. Therefore, to use ground-based interferometers, it would be required to minimize the missing sky coverage by combining observations from several apparatuses at different locations.
One could instead build two identical sets of interferometers on opposite sides on the Earth to obtain the full-sky coverage. In order to detect at least the 21-cm dipole in the EDGES/SARAS integration time windows, about 100 $\sim 1 \,{\rm m}^2$ element interferometer with redundant small-length baselines would be required. 
Another option would be using the {far side} of the Moon, such as the ``FARSIDE" suggested by \citet{Burns2021}: a 128-element radio interferometer planned to be deployed on the {far side} of the moon. This would mitigate {the impact of the Earth's ionospheric and other radio frequency interferences}. What one needs for a low-order multipole observation will be scaled-down version of FARSIDE in the baseline and the number of elements: FARSIDE aims at maximum 10 km baseline while low-order multipole observation will require smaller-scale baselines with a moderate number of baselines.


Future work related to our proposition is warranted. We will investigate the feasibility of using existing high-sensitivity radio telescopes or planned ones to measure the kinematic 21-cm dipole and quadrupole moments. For such a forecast, we will consider SKA precursors such as LOFAR, MWA and HERA and also the SKA itself. Some of these instruments are still optimized for observation of scales smaller than low-$\ell$ multipoles. In general, instruments with large-angle primary beam should be optimal, which we intend to investigate.
It is also possible that the Milky Way foreground signal will be modulated differently from the 21-cm background if we consider e.g. the seasonal variation in the motion of the Earth. We will study if we can reduce the degeneracy in estimating foreground and 21-cm background parameters from this modulation difference. We will try to find a telescope configuration optimal for the 21-cm dipole and quadrupole measurements covering the redshift range of $z\sim 6$--$30$. 


As we have observed in our paper, additional information on the low multipole 21-cm background increases the accuracy of the foreground parameters as well as the astrophysical parameters, even though the improvement in the latter is only about a few per cent.
If one can factor out the foreground from the kinematic 21-cm multipoles (at least from the dipole moment), the observed dipole spectrum can be readily compared with the monopole spectrum for consistency. Such a consistency check may well resolve the issue on the mutually inconsistent monopole spectra from the EDGES and the SARAS. As long as a successful removal of the foreground from each multipole is made, the sole dependence of the spectrum of each induced 21-cm multipole on the spectrum of the 21-cm monopole adds redundancy in determining the monopole spectrum, under the restriction that higher the multipole-order ($\ell$) is, the more difficult the observation becomes due to the decrease of the signal with the factor $\beta^\ell$. In addition, to achieve such a useful redundancy, intrinsic foreground multipole spectra better be separately measured with high accuracy as demonstrated in Fig.~\ref{fig:FIG6}. Overall, we find that improved or new observations in the foreground, the CMB monopole temperature, and low-$\ell$ multipoles have the potential in constraining cosmological and astrophysical parameters with higher accuracy. 

\begin{acknowledgments}
Authors thank Jens Chluba, Donghui Jeong, Marc Kamionkowski and Jordan Mirocha for useful discussions and Jens Chluba for comments on the manuscript. SCH is supported by the Horizon Fellowship at Johns Hopkins University. KA is supported by NRF-2021R1A2C1095136 and RS-2022-00197685.
\end{acknowledgments}

\bibliographystyle{aasjournal}
\bibliography{refs_editable}

\appendix
\restartappendixnumbering

We provide a table (Table~\ref{tab:coefficients}) listing the fitting coefficients for the measured foreground multipole spectra up to $\ell=3$. For this, we used the code PyGDSM to extract multipoles at uniformly spaced, 150 frequency bins in the frequency range [50, 200] MHz based on the LFSS data.

\begin{deluxetable}{cccccccc}[h]
\label{tab:coefficients}
\tablecaption{Coefficients $a_{\ell m,i}$ in the 4-th order log-polynomial fitting for the foreground temperature anisotropy ${a'}_{\ell m}(\nu')$ }
\tablehead{$\ell$ & $m$ & Type & $a_{\ell m,0}$ & $a_{\ell m,1}$ & $a_{\ell m,2}$ & $a_{\ell m,3}$ & $a_{\ell m,4}$}
\startdata
0 & 0 & R & 9450.9 & -23166 & 25046 & -13500 & 2873.4 \\
1 & 0 & R & -551.63 & 1372.4 & -1666.0 & 1100.4 & -302.21 \\
1 & 1 & R & 644.51 & -1609.5 & 1958.0 & -1295.4 & 356.90 \\
1 & 1 & I & 1975.3 & -4822.5 & 5459.1 & -3232.8 & 780.62 \\
2 & 0 & R & -1634.5 & 3980.0 & -4615.9 & 2859.1 & -726.65 \\
2 & 1 & R & -1169.3 & 2813.1 & -3297.1 & 2092.4 & -542.59 \\
2 & 1 & I & 208.17 & -505.43 & 603.75 & -393.37 & 105.36 \\
2 & 2 & R & -1563.4 & 3813.1 & -4391.0 & 2683.6 & -672.43 \\
2 & 2 & I & 131.98 & -330.35 & 424.85 & -304.45 & 90.071 \\
3 & 0 & R & 321.19 & -798.49 & 969.19 & -640.29 & 175.81 \\
3 & 1 & R & 29.333 & -62.945 & 50.858 & -11.005 & -3.9156 \\
3 & 1 & I & -662.79 & 1601.3 & -1756.2 & 984.41 & -219.44 \\
3 & 2 & R & -67.008 & 161.19 & -187.36 & 117.26 & -29.942 \\
3 & 2 & I & -536.32 & 1277.1 & -1380.0 & 757.27 & -161.73 \\
3 & 3 & R & -110.76 & 277.03 & -332.55 & 215.24 & -58.096 \\
3 & 3 & I & -620.36 & 1481.5 & -1574.8 & 832.73 & -168.00 
\enddata
\tablecomments{This is to fit foreground ${a'}_{\ell m}$ measured in the reference frame $\hat{z}=\hat{\beta}_\odot^{\rm CMB}$, and is inclusive of the Milky Way and the extragalactic contribution. Each multipole moment is fitted by ${a'}_{\ell m}(\nu')=\sum_{i} a_i [\ln (\nu/\nu_0)]^i$. Coefficients for the real ("R") and imaginary("I") parts of each multipole are separately shown.}
\end{deluxetable}

We also provide the conjugated Wigner-D matrix $D^{\ell*}_{m' m}(\hat{\beta}_{\odot,\rm MW}\rightarrow\hat{\beta}_{\odot,\rm CMB})$ for $\ell=\{1,\,2\}$. Each of these frames, with $\hat{z}=\hat{\beta}_{\odot,\rm MW}$ and $\hat{z}=\hat{\beta}_{\odot,\rm CMB}$, are first generated by taking the Euler angles ($\alpha$, $\beta$, $\gamma$)=($l$, $\pi/2-b$, $0$) in rotating the galactic axes ($\hat{R}$, $\hat{\phi}$, $\hat{Z}$) in the $z$-$y$-$z$ rotation convention, with the galactic longitude $l$ and the galactic latitude $b$ of the velocity vector $\hat{\beta}_{\odot}$. This convention make both frames (with $\hat{z}=\hat{\beta}_{\odot,\rm MW}$ and $\hat{z}=\hat{\beta}_{\odot,\rm CMB}$) have y-axes still lying on the galactic plane. We obtain
\begin{equation}
D_{m' m}^{1*} = 
\begin{bsmallmatrix}
-0.17943+0.019014 i & -0.54274+0.034442 i & -0.81938+0.017352 i \\
& & \\
-0.54335+0.022943 i & -0.63914            &  0.54335+0.022943 i  \\
& & \\
-0.81938-0.017352 i &  0.54274+0.034442 i & -0.17943-0.019014 i
\end{bsmallmatrix}
.
\end{equation}

and

\begin{equation}
D_{m' m}^{2*}=
\begin{bsmallmatrix}
0.031833-0.0068231i &  0.13679-0.023333i  &  0.35932-0.045788i &  0.62807-0.053229i  &  0.67109 -0.028436i \\
&&&&\\
0.13726 -0.020432i  &  0.40878-0.043318i  &  0.60082-0.038128i &  0.22801-0.0048286i & -0.63019 -0.013252i \\
&&&&\\
0.36093 -0.030535i  &  0.60149-0.025398i  &  0.11274           & -0.60149-0.025398i  &  0.36093 +0.030535i \\
&&&&\\
0.63019 -0.013252i  &  0.22801+0.0048286i & -0.60082-0.038128i &  0.40878+0.043318i  & -0.13726 -0.020432i \\
&&&&\\
0.67109 +0.028436i  & -0.62807-0.053229i  &  0.35932+0.045788i & -0.13679-0.023333i  & 0.031833+0.0068231i
\end{bsmallmatrix}
\end{equation}
Equations (1) and (2) are generated by the following group property:
\begin{equation}
D^{\ell*}_{m' m}(\hat{z}_{1}\rightarrow\hat{z}_{2}) = \sum_{m''} D^{\ell*}_{m' m''}(\hat{z}_{1}\rightarrow\hat{z}) D^{\ell*}_{m'' m}(\hat{z}\rightarrow\hat{z}_{2}),
\end{equation}
where $\hat{z}_1$ and $\hat{z}_2$ are the directions of $z$-axes, and $\hat{z}$ is that of an intermediate $z$-axis. In our case, $\hat{z}_1 = \hat{\beta}_{\odot,\rm MW}$,  $\hat{z}_2 = \hat{\beta}_{\odot,\rm CMB}$ and $\hat{z} = \hat{Z}$ giving
\begin{equation}
D^{\ell*}_{m' m}(\hat{z}_{1}\rightarrow\hat{z}_{2})=\sum_{m''} D^{\ell*}_{m' m''}(-l_{\odot, \rm MW},\,-(\pi/2-b_{\odot, \rm MW}),\,0) D^{\ell*}_{m'' m}(l_{\odot, \rm CMB},\,\pi/2-b_{\odot, \rm CMB},\,0),
\end{equation}
where the set of Euler angles ($\alpha$, $\beta$, $\gamma$) are shown as the arguments of $D^{\ell *}_{m'm}$ on the right-hand side, and the galactic longitude and latitude ($l$, $b$) of the solar velocities are now with corresponding subscripts.


\end{document}